\documentclass[twocolumn,pre,nobibnotes,altaffilletter,amsmath,amssymb,amsfonts]{revtex4}
\usepackage[latin1]{inputenc}
\usepackage{graphicx}
\usepackage{amsmath}
\usepackage{latexsym}
\usepackage{epsfig}

\usepackage[latin1]{inputenc}
\usepackage{graphicx}
\usepackage{amsmath}
\usepackage{latexsym}
\usepackage{epsfig}

\begin{document}
\title{Strongly correlating liquids and their isomorphs}

\author{Ulf R. Pedersen}
\affiliation{Department of Chemistry, University of California, Berkeley, California 94720, USA}
\author{Nicoletta Gnan, Nicholas P. Bailey, Thomas B. Schr{\o}der, and Jeppe C. Dyre}
\affiliation{DNRF Center ``Glass and Time'', IMFUFA, Dept. of Sciences, Roskilde University, P.O. Box 260, DK-4000 Roskilde, Denmark}

\date{\today}
\begin{abstract}
This paper summarizes the properties of strongly correlating liquids, i.e., liquids with strong correlations between virial and potential energy equilibrium fluctuations at constant volume. We proceed to focus on the experimental predictions for strongly correlating glass-forming liquids. These predictions include i) density scaling, ii) isochronal superposition, iii) that there is a single function from which all frequency-dependent viscoelastic response functions may be calculated, iv) that strongly correlating liquids are approximately single-parameter liquids with close to unity Prigogine-Defay ratio, and v) that the fictive temperature initially decreases for an isobaric temperature up jump. The ``isomorph filter'', which allows one to test for universality of theories for the non-Arrhenius temperature dependence of the relaxation time, is also briefly discussed.
\end{abstract}

\maketitle

\section{Introduction}

After the initial reports in early 2008 of the existence of a class of strongly correlating liquids  \cite{ped08a,ped08b}, these liquids were described in four comprehensive publications that appeared  later in 2008 and in 2009 in the Journal of Chemical Physics \cite{I,II,III,IV}. This paper briefly summarizes the properties and characteristics of strongly correlating liquids as detailed in Refs. \onlinecite{I,II,III,IV} and present a number of new computer simulations. We list a number of experimental predictions for strongly correlating liquids, focusing on glass-forming liquids  since this volume constitutes the proceedings of the Rome conference held in September 2009 (6IDMRCS). The main message is that the class of strongly correlating liquids, which includes the van der Waals and metallic liquids, are simpler than liquids in general. This explains, for instance, the long known observation that hydrogen-bonded liquids have several peculiar properties.

\section{Strong virial / potential energy correlations in liquids}

Consider a system of $N$ particles in volume $V$ at temperature $T$. The virial $W$ is defined by writing the pressure $p$ is a sum of the ideal gas term $Nk_BT/V$ and a term reflecting the interactions as follows

\begin{equation}\label{Wdef}
pV \,=\,
Nk_BT+W\,.
\end{equation}
If $U({\bf r}_1,..., {\bf r}_N)$ is the potential energy function, the virial, which has dimension of energy, is given \cite{lan80,all87,cha87,rei98,han05} by

\begin{equation}\label{Wexp}
W({\bf r}_1,..., {\bf r}_N)\,=\, 
-1/3 \sum_i {\bf r}_i \cdot {\bf \nabla}_{{\bf r}_i} U({\bf r}_1,..., {\bf r}_N)\,.
\end{equation}
Equation (\ref{Wdef}) describes thermodynamic averages, but it also applies for the instantaneous values if the virial is defined by Eq. (\ref{Wexp}) and the temperature is defined from the kinetic energy in the usual fashion \cite{lan80,all87,cha87,rei98,han05}. 

If $\Delta U$ is the instantaneous potential energy minus its average and $\Delta W$ the same for the virial at any given state point, the $WU$ correlation coefficient $R$ is defined by (where sharp brackets denote equilibrium NVT ensemble averages)

\begin{equation}\label{R}
R \,=\,
\frac{\langle\Delta W\Delta U\rangle}
{\sqrt{\langle(\Delta W)^2\rangle\langle(\Delta U)^2\rangle}}\,.
\end{equation}
By the Cauchy-Schwarz inequality the correlation coefficient obeys $-1\leq R\leq 1$. We define {\it strongly correlating liquids} by the condition $R>0.9$
\cite{I}. The correlation coefficient is state-point dependent, but for all of the several liquids we studied by simulation \cite{I,III,sch09} $R$ is either above $0.9$ in a large part of the state diagram, or not at all.

\begin{figure}
\begin{center}
\includegraphics[width=8cm]{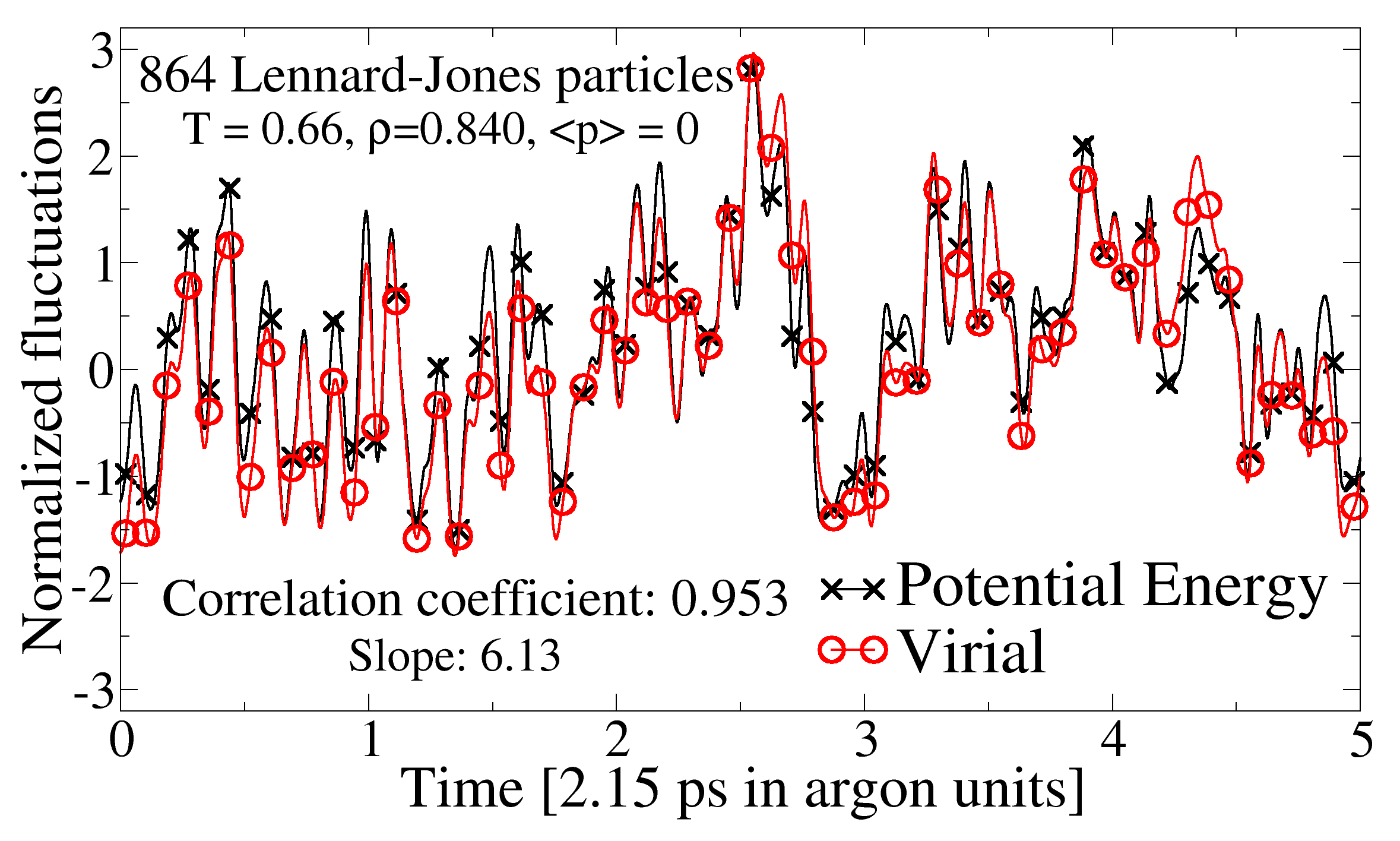}
\includegraphics[width=8cm]{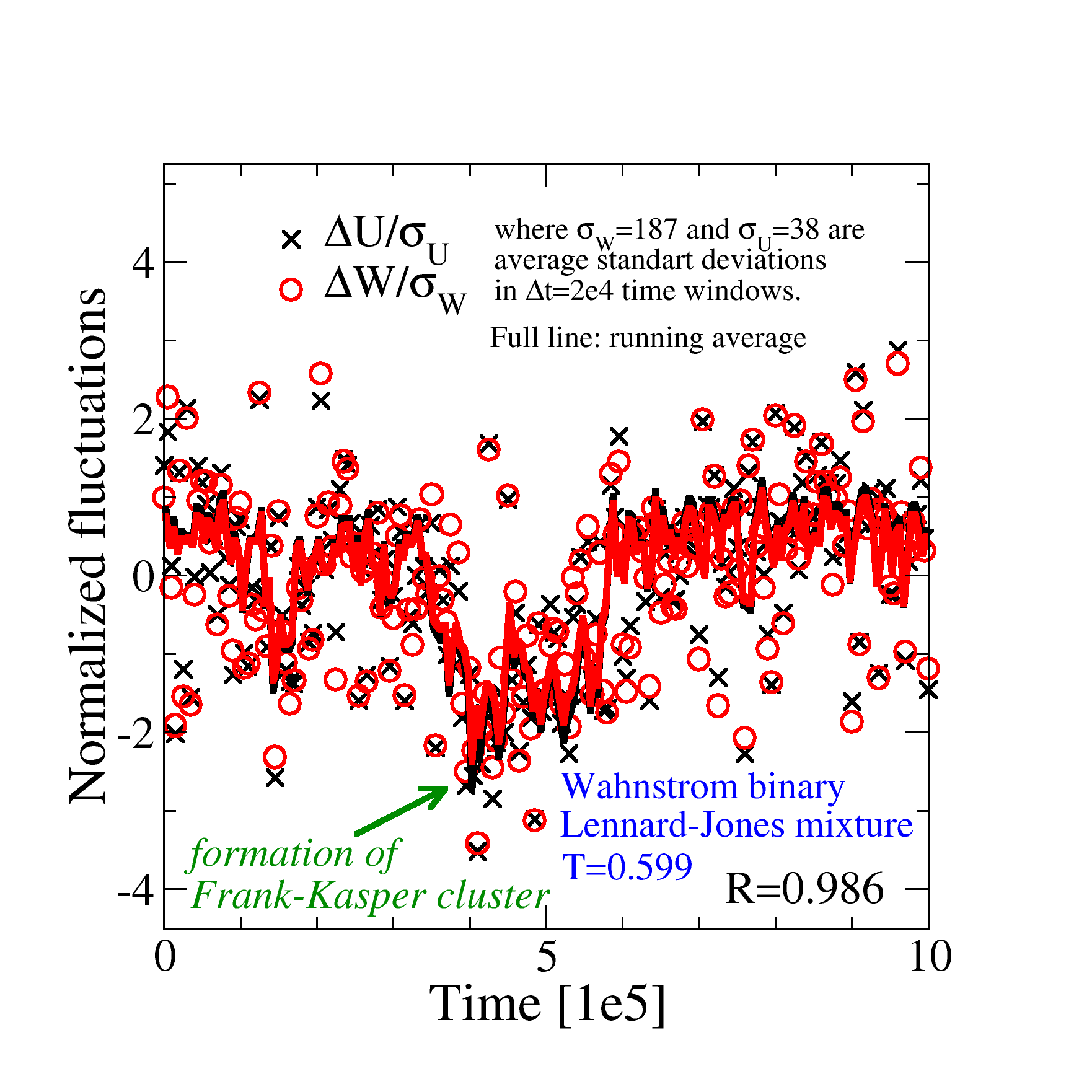}
\end{center}
\caption{(a) Instantaneous normalized equilibrium fluctuations of virial and potential energy in the standard single-component Lennard-Jones liquid at constant volume (NVT simulation). $W(t)$ and $U(t)$ correlate strongly.
(b) The same for the supercooled Wahnstr{\"o}m binary Lennard-Jones mixture \cite{wah94}; here $W(t)$ and $U(t)$ correlate strongly even during the formation of a so-called Frank-Kaspers cluster \cite{ped10}.}
\label{fig1}
\end{figure}

Figure \ref{fig1} shows two examples of constant-volume thermal equilibrium fluctuations of virial and potential energy for two model systems, the standard Lennard-Jones (LJ) liquid and the Wahnstr{\"o}m binary Lennard-Jones mixture \cite{wah94}. In both cases there are strong virial / potential energy correlations. In (b) one sees striking dips in the potential energy; these dips reflect the existence of transient clusters in the liquid characterized by the same short range order as the crystal \cite{ped10}. During the dips virial and potential energy also correlate strongly. Actually, the correlation even survives crystallization \cite{II}. Thus the property of strong virial / potential energy correlations is quite robust; even complex systems like biological membranes may exhibit strong correlations \cite{ped10a}.

\begin{figure}
\begin{center}
\includegraphics[width=8cm]{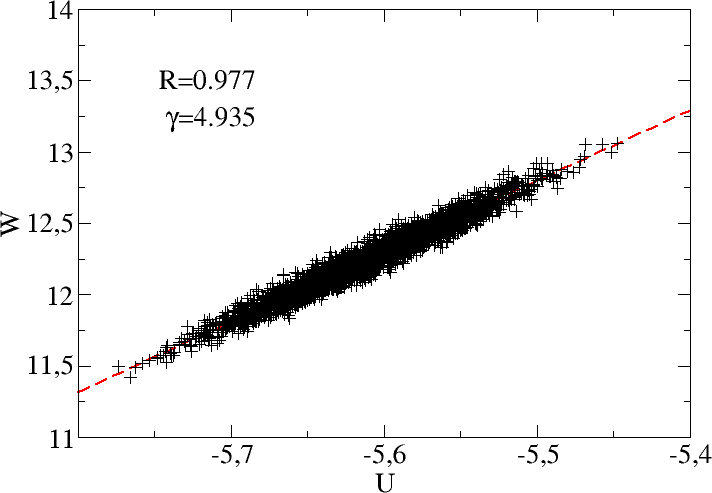}
\includegraphics[width=8cm]{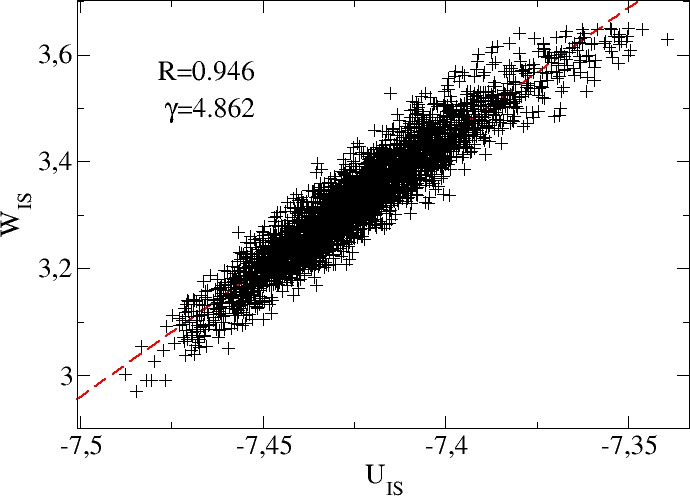}
\end{center}
\caption{(a) Virial / potential energy correlations for the Kob-Andersen binary Lennard-Jones liquid (1000 particles studied by Monte Carlo simulation, $\rho =1.264$, $T=1.24$ in standard LJ units). (b) Inherent state energies and virials of the simulation in (a); the correlation is still high. The slope $\gamma$ defined by $\Delta W(t)\cong \gamma \Delta U(t)$ is slightly different, but comparable to that of the true dynamics.}
\label{fig2}
\end{figure}

One way to illuminate the correlations is to plot instantaneous values of virial and potential energy versus one another in so-called scatter plots. Figure \ref{fig2}(a) shows an example of this with data taken from a simulation of the Kob-Andersen binary Lennard-Jones (KABLJ) liquid \cite{kob94}. This has become the standard liquid for studying viscous liquid dynamics, because it is difficult to crystallize (this requires simulating for more than 100 microseconds (Argon units)\cite{tox09}). The ``slope'' $\gamma$ of the scatter plot gives the proportionality constant of the fluctuations according to

\begin{equation}\label{gamma}
\Delta W(t)\,\cong\, \gamma \Delta U(t)\,.
\end{equation}
The number $\gamma$, which varies slightly with state point, is roughly 6 for the standard LJ liquid, roughly 5 for the KABLJ liquid, and roughly 8 for the OTP model studied below in Fig. \ref{fig7}. 

Since viscous liquid dynamics consists of long-time vibrations around potential energy minima -- the so-called inherent states \cite{sti83} -- followed by rapid transitions between the inherent states \cite{gol69,sch00}, it is interesting to study the inherent dynamics analogue of Fig. \ref{fig2}(a). This is done in Fig. \ref{fig2}(b), which gives the same simulation data after minimizing the configurations' potential energy using the conjugate gradient method. The correlations are still present and the ``slope'' $\gamma$ doesn't change very much -- even though the virial decreased by more than 60\% going from (a) to (b). This confirms the robustness of virial / potential energy correlations.

\begin{figure}
\begin{center}
\includegraphics[width=9cm]{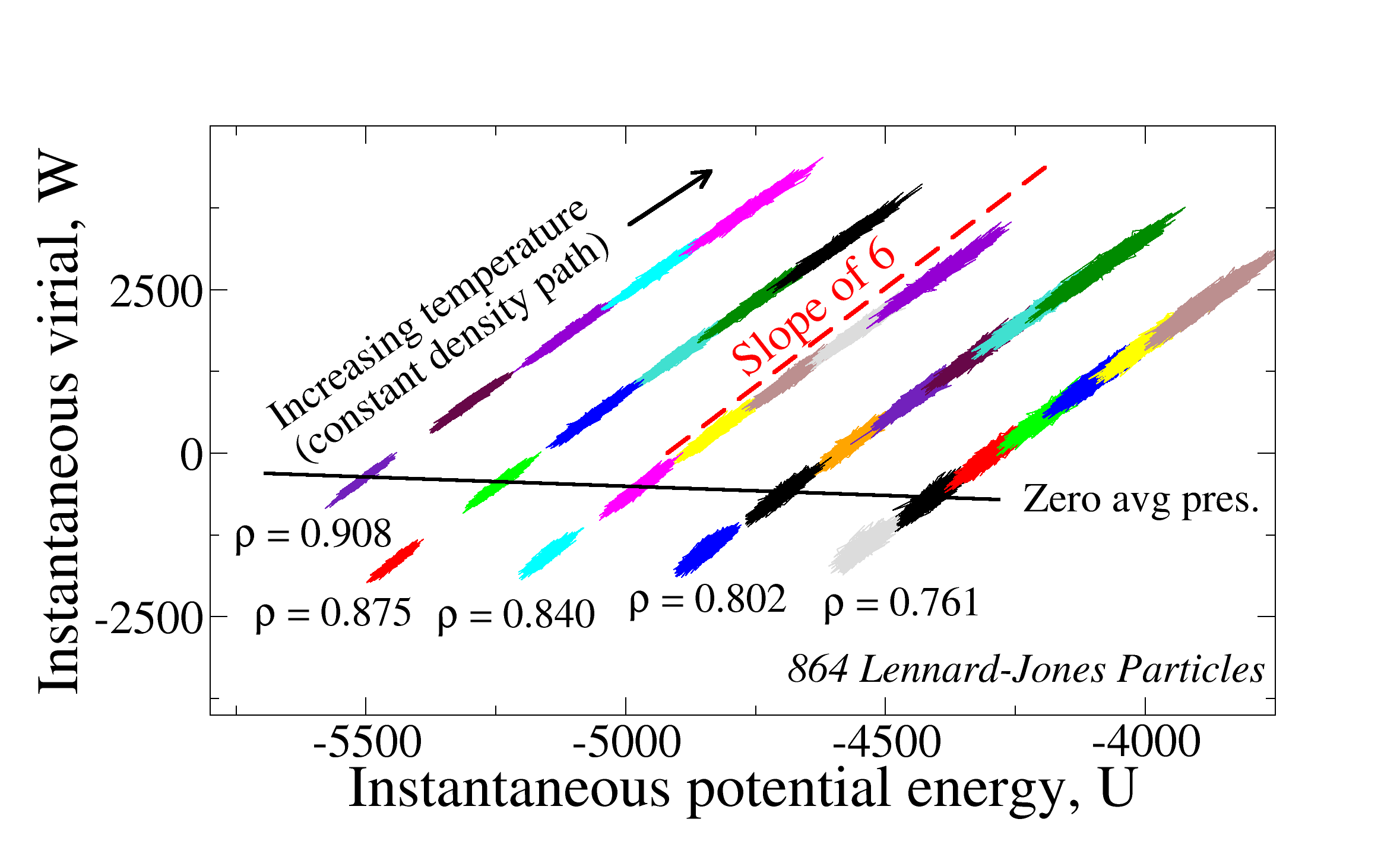}
\includegraphics[width=9cm]{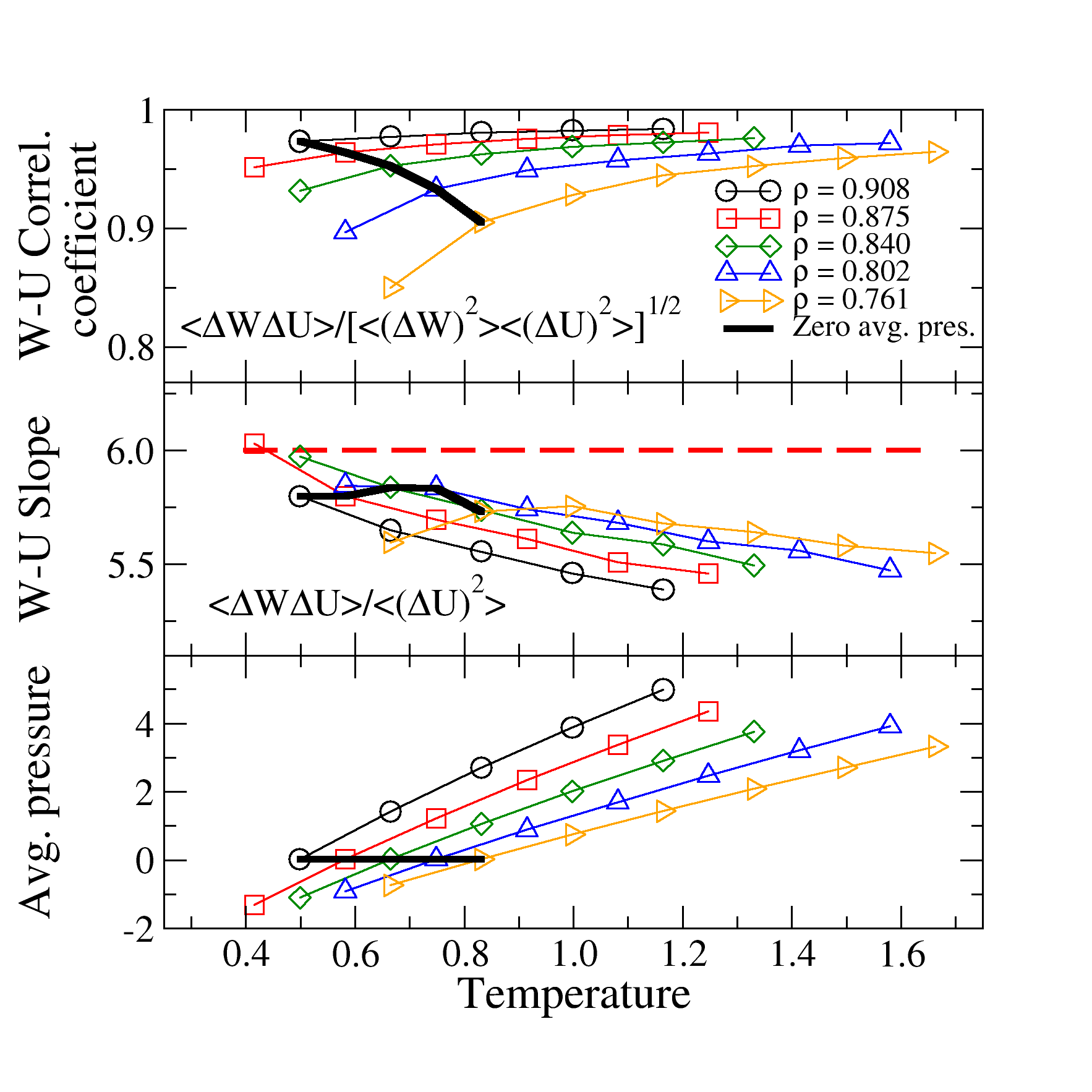}
\end{center}
\caption{(a) Scatter plot of the $WU$ thermal equilibrium fluctuations at constant volume for the standard single-component LJ liquid, and (b) plots of various quantities as functions of temperature for the different densities studied. The full black line marks state points of zero average pressure.}
\label{fig3}
\end{figure}

A convenient way to get an overview of a liquid's $WU$  thermal equilibrium fluctuations at constant volume is to collect scatter plots for several state points in a common diagram. Figure \ref{fig3} (top) shows such a plot for the standard LJ liquid. Each state point is represented by one color. As in Fig. \ref{fig2} the strong correlation is reflected in the fact that the ovals are highly elongated. For each value of the density the ovals form almost straight lines with slope close to 6 (in Ref. \onlinecite{II} it was shown that during and after constant-volume crystallization the system's scatter plots fall on the extension of the line). The bottom three figures show the correlation coefficient $R$ (Eq. (\ref{R})), the ``slope'' $\gamma$, and the average pressure as functions of temperature for the different densities. Clearly, both $R$ and $\gamma$ are somewhat state-point dependent. At a given density $R$ increases with temperature whereas $\gamma$ decreases; at a given temperature $R$ increase with increasing density. The thick black lines mark state points of zero average pressure. Note that the density effect of increasing $R$ ``wins'' over the temperature effect of decreasing $R$ upon cooling at constant low pressure. Thus one expects higher correlations upon supercooling a liquid, which is an important observation when it comes to focusing on glass-forming liquids.

How common are strong $WU$ correlations? In Ref. \onlinecite{I} we reported simulations of 13 different model liquids. All liquids with van-der-Waals type interactions were found to be strongly correlating ($R>0.9$), whereas models of the two hydrogen-bonding liquids water and methanol were not. Although much remains to be done by means of theory and simulation, it has now been established without reasonable doubt that liquids can be classified into two classes: (i) The class of strongly correlating liquids, which includes the van der Waals and metallic liquids; this liquid class has a number of regularities and simple properties. (b) All remaining liquids -- the hydrogen-bonded, the covalently bonded, and (strongly) ionic liquids -- which are much more complicated.

\section{Cause of strong virial / potential energy correlations}

\begin{figure}
\begin{center}
\includegraphics[width=8cm]{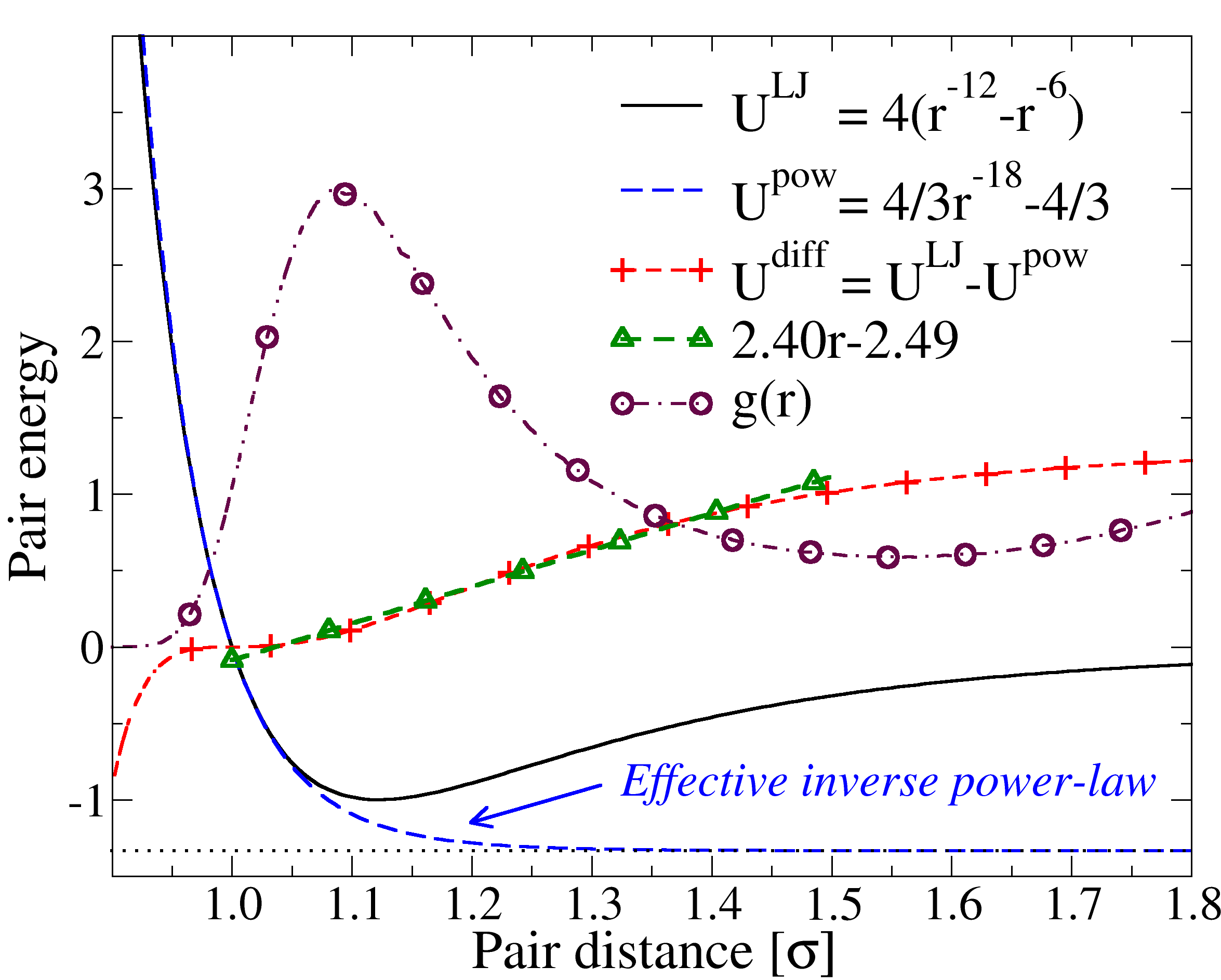}
\includegraphics[width=8cm]{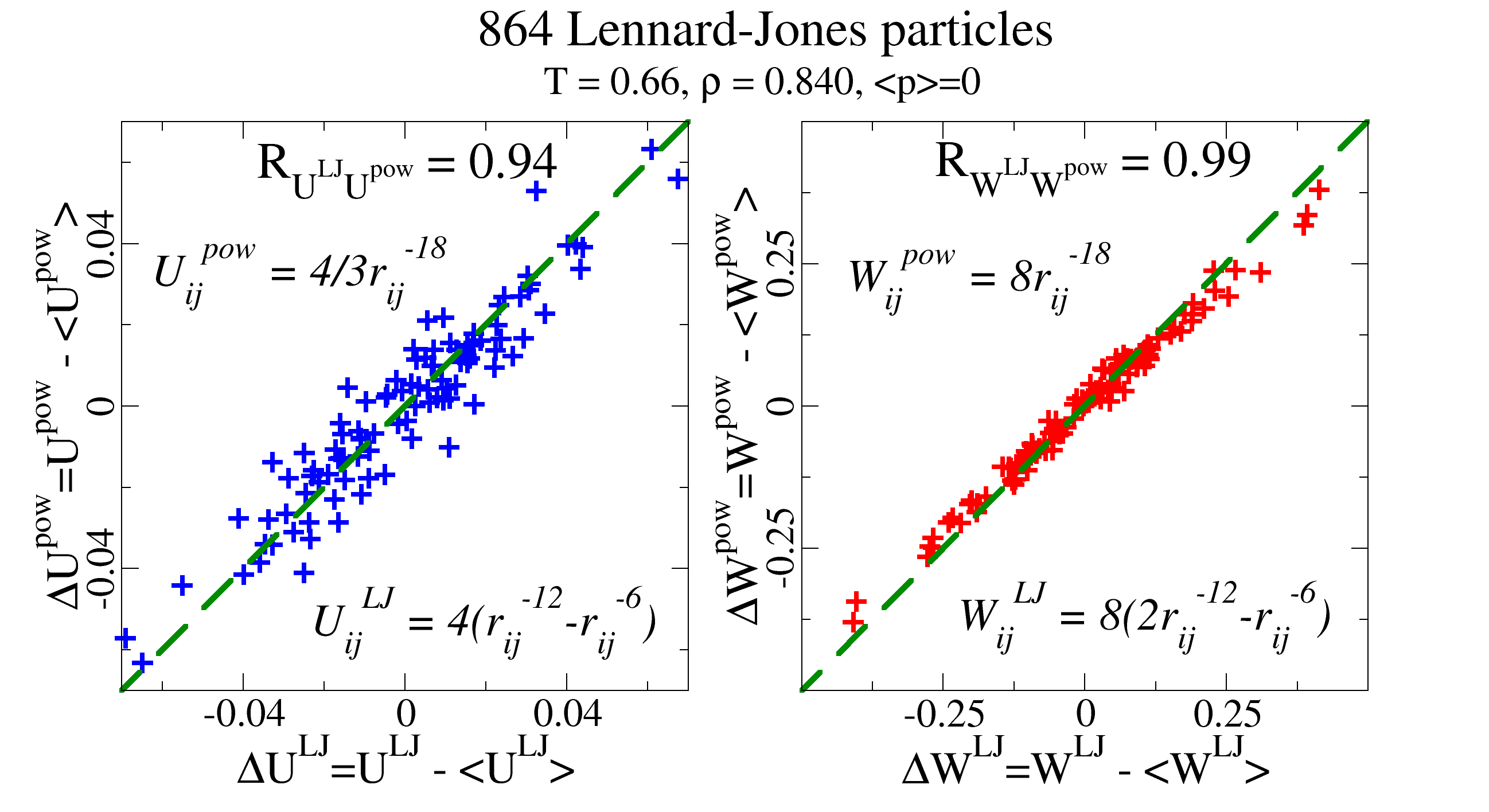}
\end{center}
\caption{(a) Approximation of the LJ potential by an effective inverse power law (IPL) potential $\propto r^{-18}$. The blue dotted curve marks the IPL potential, which approximates the LJ potential well at small interparticle spacing. The red open circles mark the radial distribution function at a typical low-pressure state point. The difference between the LJ potential and the IPL potential is approximately linear in $r$; this fact forms the basis for the ``extended inverse power law'' (eIPL) approximation (Eq. (\ref{eIPL})) \cite{II,III}. (b) Two figures demonstrating that the LJ potential and its virial in their thermal equilibrium fluctuations correlate strongly to the same quantities for the $r^{-18}$ IPL potential.}
\label{fig4}
\end{figure}

Before discussing the consequences of strong virial / potential energy correlations we briefly reflect on the cause of the correlations. The starting point is the well-known fact \cite{lan80,all87,cha87,rei98,han05} that for any liquid in which the particles (of one or more types) interact with purely repulsive inverse power-law forces, $v(r)\propto r^{-n}$, there is 100\%  correlation between $W$ and $U$: $W(t)= \gamma U(t)$ where 

\begin{equation}\label{gammaIPL}
\gamma\,=\,
\frac{n}{3}\,.
\end{equation}
From the values of $\gamma$ close to 6 observed for the LJ liquid one would expect that, if the LJ liquid somehow corresponds to an IPL liquid, the exponent $n$ is close to 18. Although at first sight this may seem strange given the $r^{-6}$ and $r^{-12}$ terms that enter into the definition of the LJ potential, a potential proportional to $r^{-18}$ does indeed give a good fit to the repulsive part of the LJ potential (Fig. \ref{fig4} (a)). The reason that a much larger exponent than 12 is required is that the attractive $r^{-6}$ term makes the LJ repulsion much steeper than that of the $r^{-12}$ term alone. Figure \ref{fig4} (b) shows that both potential energy and virial fluctuations of the LJ liquid are well represented by those of an  $r^{-18}$ IPL potential.

In our first publications on strongly correlating liquids \cite{ped08a,ped08b} it was suggested that the strong correlations derive from particle close encounters taking the intermolecular distance to values below the LJ potential minimum, at which the IPL potential is a good approximation. It quickly became clear, however, that this is not the full explanation; thus this can explain neither the existence of strong correlations in the crystal (above 99\% at low temperatures \cite{II}), nor the existence of correlations at low pressures at which nearest-neighbor interparticle distances fluctuate around the LJ potential's minimum distance. Also, the original explanation is a single-pair explanation, which would imply that the strong correlations should be present as well in constant pressure ensembles. This contradicts our finding that switching from constant volume to constant pressure reduces $R$ from values above 0.9 to values around 0.1 \cite{III}. 

References \onlinecite{II} and  \onlinecite{III} detail the more complete explanation of the cause of strong correlations. The difference between the IPL potential and the LJ potential is plotted in Fig. \ref{fig4} (a) as the red dashed curve. The green dashed curve is a straight line, which approximates the red dashed curve well around the LJ minimum (i.e., over the entire first peak of the structure factor). Thus over the most important intermolecular distances an ``extended'' inverse power law potential, eIPL, defined by

\begin{equation}\label{eIPL}
v_\textrm{eIPL}(r) \,=\,
A r^{-n}+B+Cr\,
\end{equation}
gives a good approximation to the LJ potential, $v_\textrm{LJ}(r)\cong v_\textrm{eIPL}(r)$. It has been shown by simulation that the linear ``quark confining'' term of the eIPL potential gives a contribution to the total potential energy that fluctuates little {\it at constant volume} \cite{II,III}. Thus as regards {\it fluctuations}, the pure IPL gives representative results. This explains why the IPL approximation works so well and why the strong correlations disappear when going to constant pressure ensembles. This also explains why several IPL liquid properties are not shared by LJ-type liquids (e.g., the IPL equation of state is generally quite wrong and does not allow for low-pressure stable liquid states, and the IPL free energy and bulk modulus are quite wrong). 

While the eIPL approximation explanation of strong $WU$ correlations for physically realistic cases, there are also strong correlations in the purely repulsive Weeks-Chandler-Andersen \cite{wee71} version of the KABLJ liquid \cite{kob94,ber09,cos09a}. The slope $\gamma$ here varies quite a lot  (from 5.0 to 7.5) over the range of densities and temperatures in which $\gamma$ is fairly constant for the KABLJ liquid. Our simulations show that the strong correlations for the WCA case is a single-particle-pair effect, not the cooperative effect that only applies at constant volume conditions, observed for LJ-type liquids. More work is needed to illuminate the correlation properties of this interesting (but physically unrealistic) potential.

\section{Isomorphs: Curves of invariance in the phase diagram}

This section defines isomorphs and summarizes their invariants. As shown in Ref. \onlinecite{IV} a liquid has isomorphs if and only if the liquid is strongly correlating. An isomorph is a curve in the phase diagram along which a number of properties are invariant. 

For any microscopic configuration $({\bf r}_1,\, ...\, , {\bf r}_N)$ of a thermodynamic state point with density $\rho$, the ``reduced'' coordinates are defined by $\tilde{\bf r}_i\equiv\rho^{1/3} {\bf r}_i$. State points (1) and (2) with temperatures $T_1$ and $T_2$ and densities $\rho_1$ and $\rho_2$ are said to be {\it isomorphic} \cite{IV} if, whenever two microscopic configurations $({\bf r}_1^{(1)},\, ...\, , {\bf r}_N^{(1)})$ and $({\bf r}_1^{(2)},\, ...\, , {\bf r}_N^{(2)})$ have identical reduced coordinates, to a good approximation they have proportional configurational NVT Boltzmann probability factors:

\begin{equation}\label{isodef}
e^{-U({\bf r}_1^{(1)},\, ...\, , {\bf r}_N^{(1)})/k_BT_1}\, =\, C_{12}\,e^{-U({\bf r}_1^{(2)},\, ...\, , {\bf r}_N^{(2)})/k_BT_2}\,.
\end{equation}
The constant $C_{12}$ here depends only on the state points (1) and (2), not on the microscopic configurations. {\it Isomorphic curves} in the state diagram are defined as curves for which any two state points are isomorphic. The property of having isomorphs is generally approximate -- only IPL liquids have exact isomorphs. For this reason Eq. (\ref{isodef}) should be understood as obeyed to a good approximation for the physically relevant configurations, i.e., those that do not have negligible canonical probabilities \cite{IV}.

\begin{figure}
\begin{center}
\includegraphics[width=8cm]{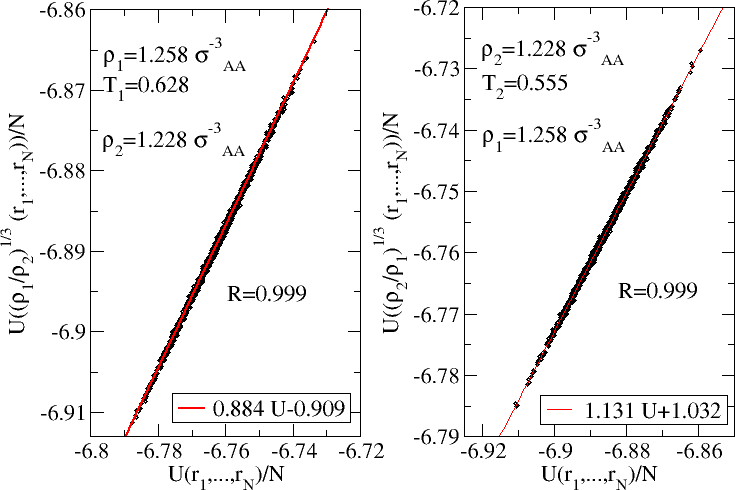}
\includegraphics[width=8cm]{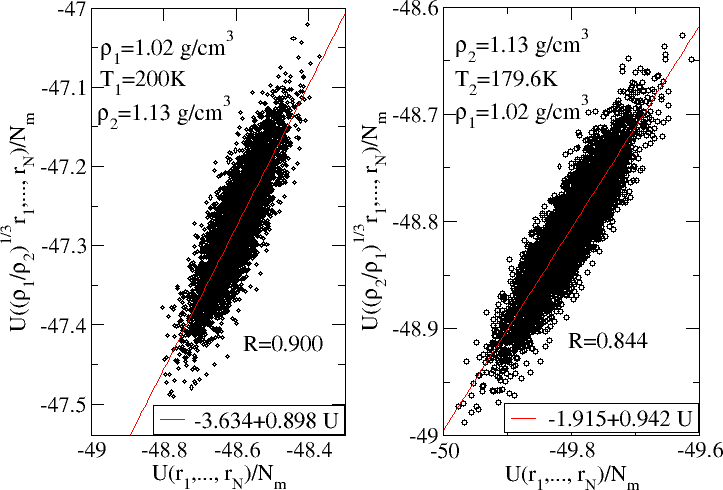}
\end{center}
\caption{Direct check of the isomorph condition for the KABLJ liquid (8000 particles), which is strongly correlating (a), and for the SPC water model (5120 molecules), which is not (b). For both liquids the consistency of the isomorph condition is checked by jumping from one  to a different density and back. This works well for the KABLJ liquid but not for SPC water; details are given in the text.}
\label{fig5}
\end{figure}

Figure \ref{fig5} illustrates Eq. (\ref{isodef}) by checking the logarithm of this equation, where (a) gives simulation data for the KABLJ liquid. We consider a number of configurations of the state point with density and temperature $(\rho_1,T_1)=(1.258, 0.628)$ in standard LJ units. For these configurations the total potential energy was evaluated. In order to investigate whether the state point has an isomorphic state point at density $\rho_2=1.228$ we scaled the simulated configurations of  the first state point to density $\rho_2$. For the scaled configurations the potential energies are plotted against the original energies of state point $1$ (top figure). According to the isomorph definition Eq. (\ref{isodef}) the best fit slope gives the ratio between the temperatures of the isomorphic state points; in this way we estimate that $T_2=0.555$.

The right panel of Fig. \ref{fig5}(a) investigates the consistency of this procedure by reversing it in order to check whether the original temperature $T_1$ is arrived at. Indeed, when this is done one does find the original temperature to be $0.628$. Two things should be noted. The first is the very strong correlation between scaled configurations, as required for having good isomorphs. The second notable fact is that the best fit lines do not pass through $(0,0)$. This shows that the constant $C_{12}$ of Eq. (\ref{isodef}) is not unity, as it would be for an IPL liquid ($C_{12}$ is determined by the contribution to the partition function coming from the linear term of the eIPL, and thus $C_{12}$ reflects the deviation from true IPL behavior).

Figure \ref{fig5}(b) makes this ``direct isomorph test'' for the non-strongly correlating liquid SPC water, starting from temperature $T_1=200$ K. From the slope of the left panel we find $T_2=179.6$ K. When the reverse jump is performed, however, one does not come back to the initial state point, but to a predicted temperature of $166.36$ K. This shows that water does not have isomorphs, consistent with the fact that it is not a strongly correlating liquid. 

\begin{figure}
\begin{center}
\includegraphics[width=8cm]{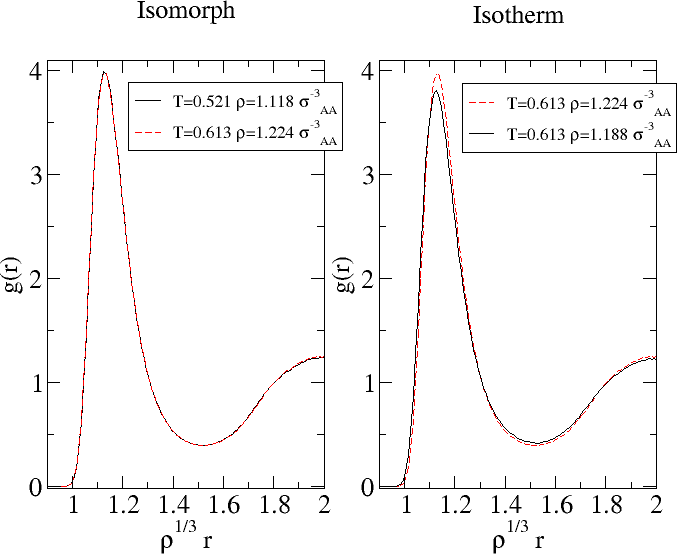}
\includegraphics[width=8cm]{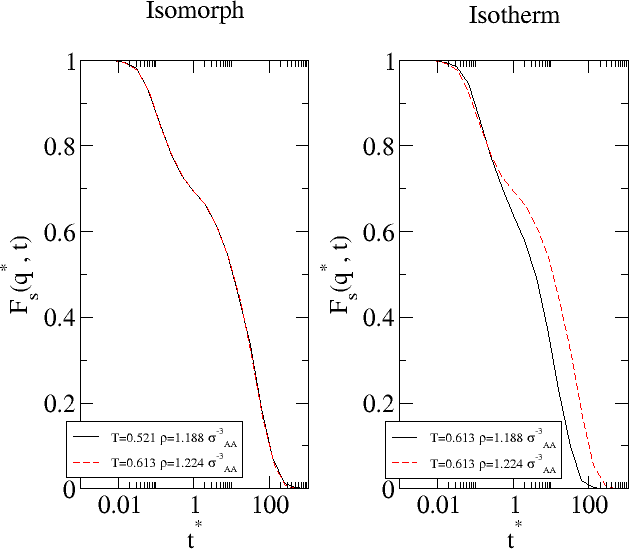}
\end{center}
\caption{(a) AA particle radial distribution function of the KABLJ liquid for two isomorphic state points (left) and for the same temperature at the two densities (right). The isomorphic state points have the same radial distribution functions. (b) The same for the AA incoherent intermediate scattering function at the k-vector corresponding to the first peak of the radial distribution function. The two isomorphic state points have the same dynamics (in reduced units, as used here).}
\label{fig4a}
\end{figure}

For the practical identification of an isomorph in the phase diagram the above method may be used. Alternatively, it has been shown \cite{IV} that to a good approximation isomorphs are characterized by

\begin{equation}\label{isochar}
\frac{\rho^\gamma}{T}\,=\,{\rm Const.}
\end{equation}
Here $\gamma$ is the above discussed ``slope'' characterized by  $\Delta W(t)\cong \gamma \Delta U(t)$. As shown in Ref. \onlinecite{IV} this quantity may be calculated to a good approximation from equilibrium fluctuations via the expression (giving the least-squared linear-regression best-fit slope of $WU$ scatter plots, compare Appendix B of Ref. \onlinecite{I})

\begin{equation}\label{gamma_expr}
\gamma\,=\,
\frac{\langle\Delta W\Delta U\rangle }{ \langle(\Delta U)^2\rangle}\,.
\end{equation}

Several physical quantities are invariant along a strongly correlating liquid's isomorphs to a good approximation. These include: 1) Thermodynamic properties like the excess entropy (i.e., in excess of the ideal gas entropy at same density and temperature) and the excess isochoric specific heat, 2) static averages like radial distribution function(s) in reduced coordinates, 3) dynamic quantities like the reduced diffusion constant, viscosity, and heat conductivity, time-autocorrelation functions in properly reduced units, average reduced relaxation times, etc. 

Figure \ref{fig4a} shows results of simulations of the KABLJ liquid at two isomorphic state points (left subfigures) and isothermal state points (right subfigures) of the AA particle radial distribution functions and the AA incoherent intermediate scattering functions, respectively. These figures confirm the prediction that isomorphic state points have identical static distribution functions and identical dynamics.

\begin{figure}
\begin{center}
\includegraphics[width=8cm]{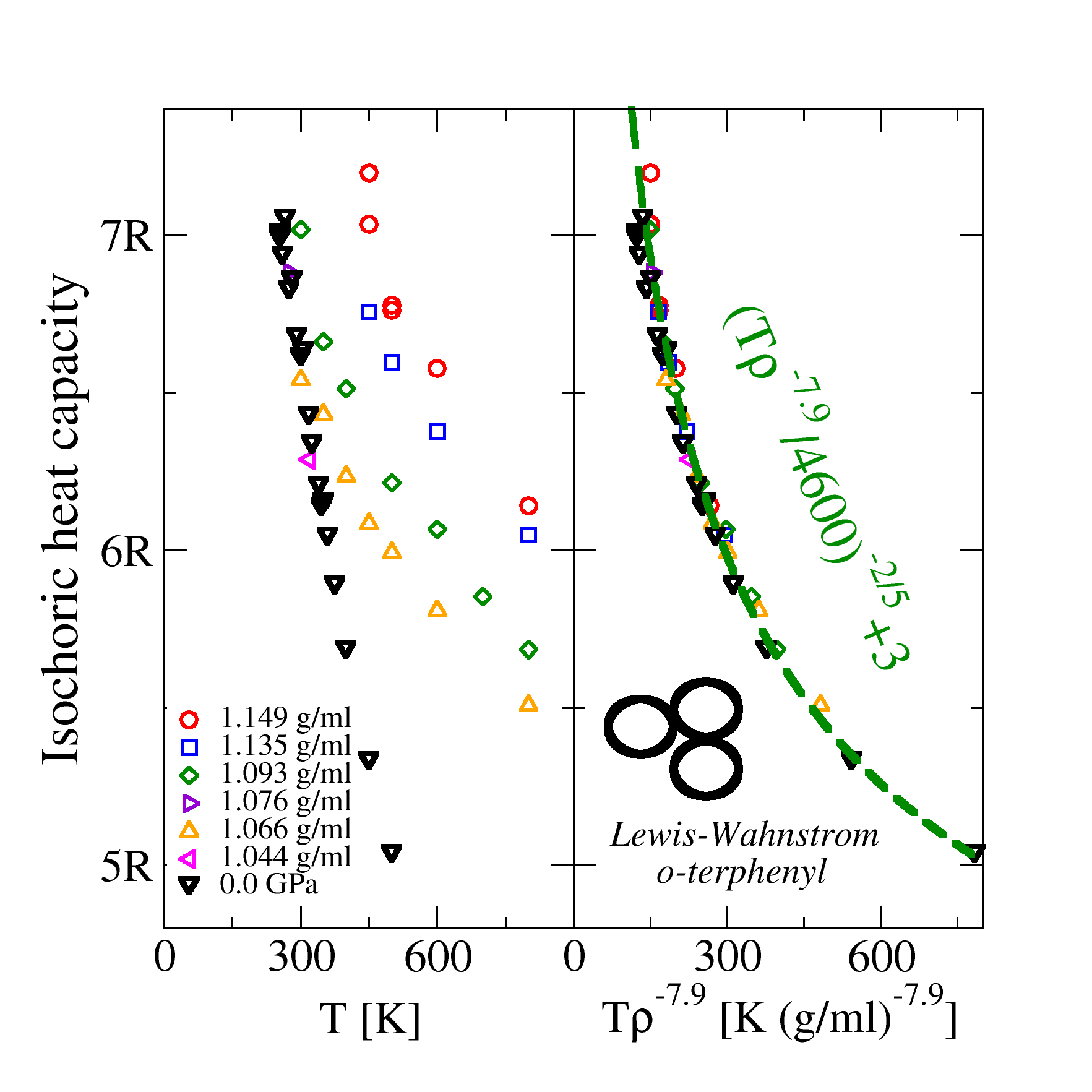}
\end{center}
\caption{$c_V$ per particle for the Lewis-Wahnstr{\"o}m model \cite{lw} of ortho-terphenyl consisting of three LJ spheres arranged with fixed bond lengths. The left panel shows the raw simulation data, The right panel shows the same data replotted as function of $\rho^{7.9}/T$ in which the exponent $7.9$ was determined from the proportionality between equilibrium virial and potential energy fluctuations \cite{sch09}. This model follows the Rosenfeld-Tarazona prediction of $c_V$ varying with temperature as $T^{-2/5}$ \cite{ros98}.}
\label{fig7}
\end{figure}

Since the isochoric specific heat is an isomorph invariant, this quantity should be a function of $\rho^\gamma/T$ for a strongly correlating liquid. Figure \ref{fig7} confirms this for the Lewis-Wahnstr{\"o}m OTP model consisting of three LJ spheres \cite{lw}.

\begin{figure}
\begin{center}
\includegraphics[width=8cm]{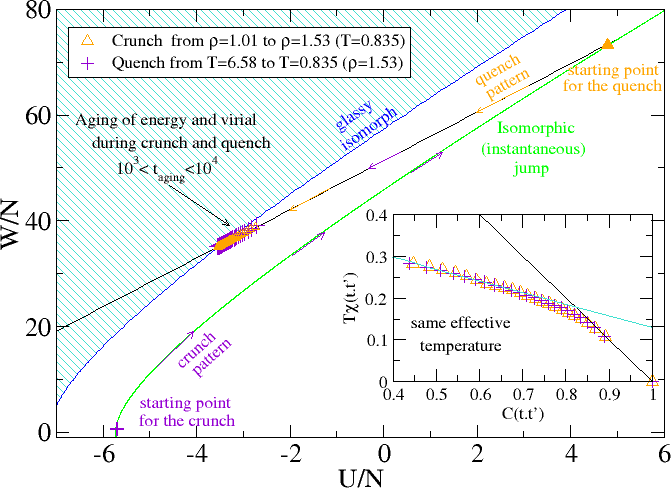}
\includegraphics[width=8cm]{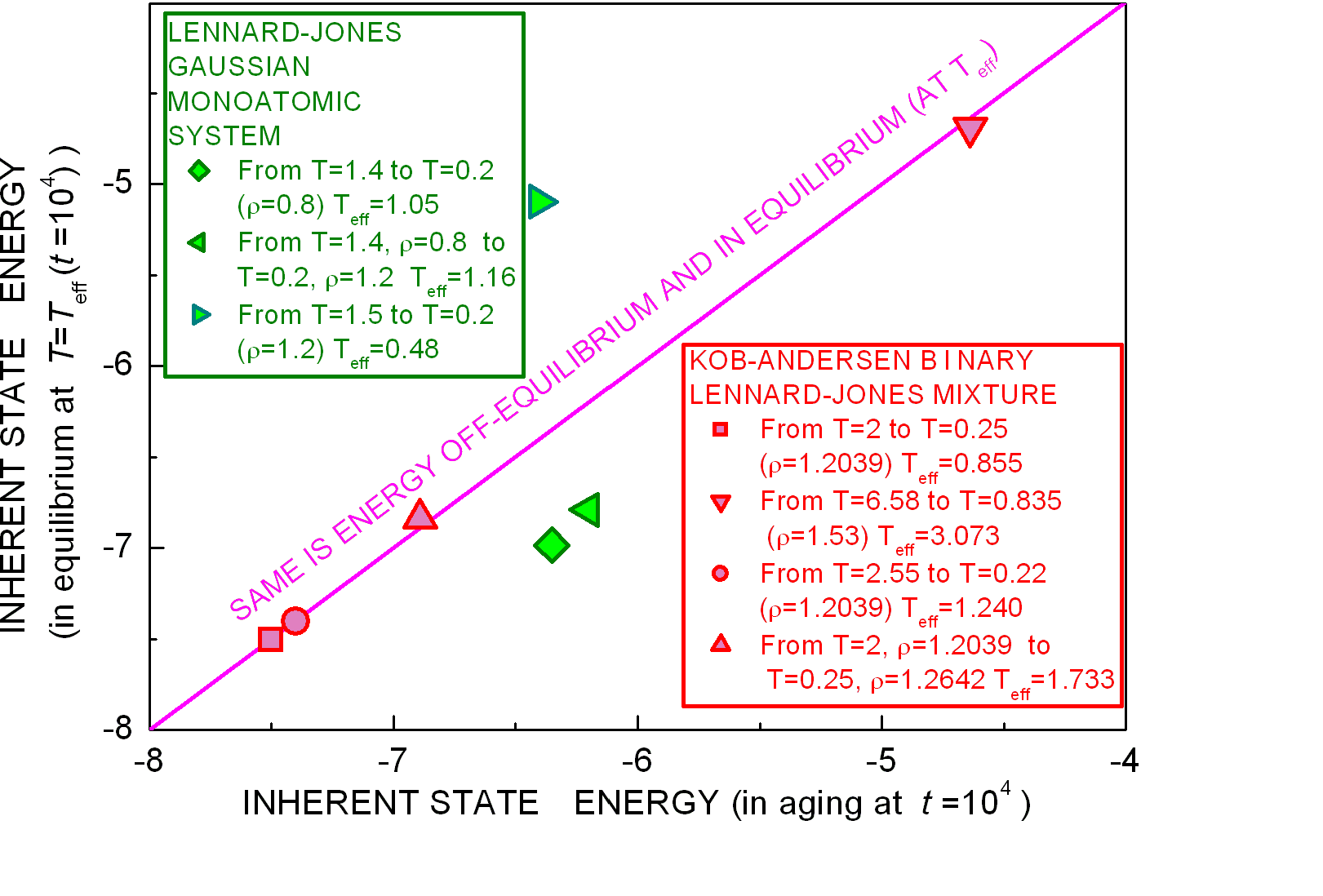}
\end{center}
\caption{(a) Virial versus potential energy for the KABLJ liquid during a temperature quench and a crunch \cite{gna10}. A crunch increases the density and keeps the temperature constant; this is equivalent to first an instantaneous jump along an isomorph (green curve) to the right density followed by a temperature quench (black curve). The inset shows that the crunch and the quench have the same fluctuation-dissipation violation factors, i.e., result in the same effective temperature. (b) 
Inherent state energies for several state points of the KABLJ liquid as the system falls out of equilibrium and freezes versus the equilibrium inherent state energy at the effective temperature (details are given in Ref. \onlinecite{gna10}).}
\label{fig6}
\end{figure}

The theory further predicts that jumps between two isomorphic state points should take the system instantaneously to equilibrium, because the Boltzmann statistical factors of two isomorphic state points by definition are proportional \cite{IV}. More generally, isomorphic state points are equivalent during any aging scheme. We recently showed that the isomorph concept can be used to throw light on the concept of an effective temperature \cite{gna10}. In particular, the theory implies that for strongly correlating liquids the effective temperature after a jump to a new (low) temperature and a new density, depends only on the new density (Fig. \ref{fig6}(a)). We showed also that this does not apply for the non-strongly correlating monatomic Lennard-Jones Gaussian liquid, confirming the general conjecture that strongly correlating liquids have simpler physics than liquids in general. Figure \ref{fig6} (b) shows a result confirming the finding of Ref. \onlinecite{gna10} that the effective temperature concept for a strongly correlating liquid makes good sense physically. On the x-axis the inherent state energies of given state points are shown as the system fell out of equilibrium. The arrested phase is characterized by an effective temperature $T_{\rm eff}$, which can be calculated in the standard way from the violation of the fluctuation-dissipation theorem \cite{gna10}. On the y-axis is shown the inherent energies found from an {\it equilibrium} simulation with temperature equal to $T_{\rm eff}$ for the corresponding arrested phase. The red points give data for the strongly correlating KABLJ liquid, the green points give data for the non-strongly correlating monatomic Lennard-Jones Gaussian liquid. Clearly, the latter system fell out of equilibrium by freezing into a part of phase space that is not characterized by $T_{\rm eff}$.

\section{The equation of state of a strongly correlating liquid}

This section shows that the Helmholtz free energy for any strongly correlating liquid is of the form

\begin{equation}\label{eqs}
F_{\rm ex} /N\,=\,
T\psi\big(T f(\rho)\big)+g(\rho)\,.
\end{equation}
For simplicity we shall not indicate ``excess'' quantities explicitly, $V$ is used as variable instead of the $\rho$, and the $N$ is ignored since it is fixed; thus we shall prove that

\begin{equation}\label{Feqs}
F \,=\,
T\psi\big(T f(V)\big)+g(V)\,.
\end{equation}
Suppose a given strongly correlating liquid's isomorphs are labeled by the variable $x$, i.e., that its isomorphs are curves of constant $x$ for which $x=x(T,V)$. Isomorphs are curves of constant (excess) entropy, as well as curves of constant (excess) $C_V$ \cite{IV}. This means that for some functions $\phi_1(x)$ and  $\phi_2(x)$ one may write

\begin{equation}\label{sandcv}
S\,=\,\phi_1(x)\,\, ,\,\,C_V\,=\,\phi_2(x)\,.
\end{equation}
Since $C_V=T(\partial S / \partial T )_V=T\phi_1'(x) ( \partial x / \partial T )_V$, if we define $m(x)\equiv\phi_2(x)/\phi'_1(x)$, one has $T (\partial x / \partial T )_V=m(x)$. Along an isochore this implies that $dx/m(x)=d\ln T$. When integrated along the isochore this gives $h(x)\equiv\int_{x_0}^x dx' / m(x')=\ln T +\alpha(V)$, implying $\exp(h(x))=Tf(V)$ where $f(V)=\exp(\alpha(V))$. In other words, lines of constant $x$ have constant $Tf(V)$. Since $x$ is merely used for labeling the isomorphs, this means that we can redefine $x$ as follows $x\equiv Tf(V)$. Integrating now 
$S=\phi_1(x)=-(\partial F/\partial T)_V$ along an isochore gives

\begin{equation}\label{F}
F\,=\, 
-\int_{T_0}^T\phi_1(x')dT'+g(V)\,=\,
-\frac{T}{x}\int_{x_0}^x\phi_1(x')dx'+g(V)\,. 
\end{equation}
The integral is some function of $x$, and we have thus derived Eq. (\ref{Feqs}).

As mentioned the slope $\gamma$ may vary with state point. The equation of state gives information about which variables $\gamma$ may depend on:

\begin{equation}\label{S}
S\,=\, 
-\left(\frac{\partial F}{\partial T}\right)_V\,=\,
-\psi\big(T f(V)\big)-T\psi'\big(T f(V)\big)f(V)\,
\end{equation}
implies that

\begin{equation}\label{E}
U\,=\, F+TS\,=\,
-T^2\psi'\big(T f(V)\big)f(V)+g(V)\,.
\end{equation}
We also have

\begin{equation}\label{p}
\frac{W}{V}\,=\, 
-\left(\frac{\partial F}{\partial V}\right)_T\,=\,
-T^2\psi'\big(T f(V)\big)f'(V)-g'(V)\,.
\end{equation}
Combining these equations leads to

\begin{equation}\label{W}
W\,=\,
V\frac{f'(V)}{f(V)}\Big(U-g(V)\Big)-Vg'(V)\,.
\end{equation}
Summarizing,  

\begin{equation}\label{W2}
W\,=\,
\gamma(V)U +h(V)\,,
\end{equation}
where

\begin{equation}\label{gammaV}
\gamma(V)\,=\,
\frac{d\ln f}{d\ln V}
\end{equation}
and 

\begin{equation}\label{h}
h(V)\,=\,-g(V)\frac{d\ln (fg)}{d\ln V}\,.
\end{equation}
Thus if $\gamma$ varies with state point, it may only depend on volume \cite{IV}. As shown in Ref. \onlinecite{IV} this result is consistent with the original experimentally based formulation of the so-called ``density scaling'' due to Alba-Simionesco and co-workers \cite{alb04}. Figure \ref{fig3}(b) shows that $\gamma$ is not rigorously constant along isochores as predicted by Eq. (\ref{W2}), although it varies only of order 10\% when temperature is tripled. This serves to emphasize that isomorphs are only approximate constructs and so are their predicted invariants (only IPL liuqids have exact isomorphs). 

The equation of state Eq. (\ref{W2}) is of the Mie-Gr{\"u}neisen form for the excess variables (excess pressure: $W/V$, and excess energy: $U$) \cite{bor54} 

\begin{equation}\label{mie}
p_{\rm ex}\,=\,\frac{W}{V}
\,=\,\frac{\gamma(V)}{V}U+\psi(V)\,.
\end{equation}
In the Mie-Gr{\"u}neisen equation of state for solids $\gamma(V)=-d\ln\omega /d\ln V$ where $\omega$ is a phonon mode eigenfrequency, $U$ is the vibrational potential energy, and $\psi(V)$ relates to the volume derivative of  the energy per atom (i.e., of the energy of the force-free configuration about which the vibrational motion occurs). Further discussion of the relation of strong $WU$ correlations to the Mie-Gr{\"u}neisen equation of state is given in Ref. \onlinecite{III}.

\section{Some experimental predictions for strongly correlating glass-forming liquids}

Strongly correlating liquids have a number of since long described properties. For instance, since the melting line in the phase diagram is an isomorph \cite{IV}, strongly correlating liquids have several invariants along their melting lines, including the radial distribution function and dimensionless transport coefficients. Such regularities have been observed in simulation and experiment; we refer to the reader to Ref. \onlinecite{IV} for more details. This section focuses on predictions for highly viscous liquids, for which the strong-correlation property implies several experimental predictions.

\begin{enumerate}

\item{{\it Density scaling}}

In the last decade, in particular since 2005, many papers appeared dealing with the so-called density scaling, which is the finding that for several glass-forming liquids the relaxation time $\tau$ at varying pressure and temperature is some function of the quantity $\rho^\gamma/T$, in which the exponent $\gamma$ is an empirical fitting parameter:

\begin{equation}\label{dens_scal}
\tau\,=\,
F(\rho^\gamma/T)\,.
\end{equation}
Neither the function $F$ nor $\gamma$ are universal. The isomorph theory \cite{IV,sch09} shows that all strongly correlating glass formers obey density scaling with the exponent $\gamma$ given by the equilibrium fluctuations at one state point (Eq. (\ref{gamma_expr})) (provided $\gamma$ is fairly constant over the relevant part of phase space). This has not yet been tested experimentally, but it is consistent with the finding that density scaling works well for van der Waals liquids \cite{rol05}, but not for hydrogen-bonded liquids \cite{gra06,rol06,leg07,rol08}. Meanwhile, density scaling has been shown to apply in computer simulations of strongly correlating liquids with $\gamma$ given by Eq. (\ref{gamma_expr}) to a good approximation  \cite{sch09,cos09}.

\item{{\it Isochronal superposition}}

Isochronal scaling is the further, fairly recent finding that for varying pressures and temperatures the dielectric loss as a function of frequency depends only on the loss-peak frequency \cite{rol03,nga05}. This is trivial if the liquid obeys time-temperature-pressure superposition (TTPS) in which case nothing changes. But many liquid do not obey TTPS, and for such liquids isochronal superposition is a new and striking regularity that works generally for van der Waals liquids, but rarely for hydrogen-bonding liquids \cite{nga05}. Since both the relaxation time and the entire relaxation spectrum are isomorph invariants, isochronal superposition must apply for any strongly correlating liquid: If temperature and pressure for two state points are such that their relaxation times are the same, the two points must belong to the same isomorph and thus have same relaxation time spectra for any observables, in particular the dielectric loss as function of frequency should be the same.

\item{{\it Frequency-dependent viscoelastic response functions}}

There are eight fundamental complex, frequency-dependent linear thermoviscoelastic response functions like, e.g., the frequency-dependent isochoric or isobaric specific heat, the frequency-dependent isobaric expansion coefficient, and the frequency-dependent adiabatic or isothermal compressibility \cite{ell07}. Standard linear irreversible thermodynamic arguments, where the Onsager relations play the role of the Maxwell relations of  usual thermodynamics, show that there are only three independent frequency-dependent response functions. If moreover stochastic dynamics is assumed as is realistic for highly viscous liquids \cite{gle98}, there are only two independent response functions \cite{mei59,gup76,ber78}. For strongly correlating liquids the further simplification appears that there is just a single independent response function \cite{ped08a,ell07,bai08c}. Since there are explicit expressions linking the different response functions (depending on the ensemble considered \cite{ell07}), this can be tested experimentally. Unfortunately it is difficult to measure thermoviscoelastic functions properly; to the best of our knowledge there are yet no reliable data for a complete set of three or more such response functions on any liquid.

\item{{\it The Prigogine-Defay ratio: Strongly correlating liquids as approximate single-parameter liquids}}

After many years of little interest the Prigogine-Defay (PD) ratio \cite{pri54,dav52,dav53} has recently again come into focus in  the scientific discussion about glass-forming liquids \cite{ell07,nie97,sch06,won07,pic08}. From a theoretical point of view the PD ratio is poorly defined since it involves extrapolations from the liquid and glass phases to a common temperature \cite{ell07,bai08c}. It is possible to overcome this problem by modifying the PD ratio by referring exclusively to linear response experiments; here the traditional difference between liquid and glass responses is replaced by a difference between low- and high-frequency values of the relevant frequency-dependent thermoviscoelastic response function \cite{ell07}. In this formulation, the property of strong virial / potential energy correlations manifests itself as a PD ratio close to unity. Actually, an extensive compilation of data showed that van der Waals bonded liquids and polymers have PD ratios close to unity \cite{urp_phd} -- even though as mentioned the traditional PD ratio is poorly defined, there are good reasons to assume that it approximates the rigorously defined ``linear'' PD ratio.

The theoretical developments of Refs. \onlinecite{ell07,ped08b,II} show that in any reasonable sense of the old concept ``single-order-parameter liquid'', strongly correlating liquids are precisely the single-order-parameter liquids. The isomorph concept makes this even more clear: State points along an isomorph have so many properties in common that they are identical from many viewpoints. In the two-dimensional phase diagram this leaves just one parameter to classify which isomorph the state point is on; thus a liquid with (good) isomorphs is an (approximate) single-parameter liquid. Note that this is consistent with the old viewpoint that single-parameter liquids should have unity PD ratio \cite{pri54,dav52,dav53}.

\item{{\it Cause of the relaxation time's non-Arrhenius temperature dependence: The isomorph filter}}

Since the relaxation time $\tau$ is an isomorph invariant for any strongly correlating liquid, any universally valid theory predicting $\tau$ to depend on some physical quantity must give $\tau$ as a function of another isomorph invariant (we do not distinguish between the relaxation time and the reduced relaxation time since their temperature dependencies are virtually identical). This gives rise to an ``isomorph filter'' \cite{IV}, showing that several well-known models cannot be generally valid. For instance, the entropy model cannot apply in the form usually used by experimentalists: $\tau\propto\exp(C/S_{\rm conf}T)$ where $C$ is a constant and $S_{\rm conf}$ is the configurational entropy; it can only be correct if $C$ varies with density as $C\propto\rho^\gamma$. Likewise, the free volume model does not survive the isomorph filter. If the characteristic volume $V_c$ of the shoving model (predicting that $\tau\propto\exp (V_c G_{\infty}/k_BT)$ \cite{dyr96,dyr06,tor09}) varies with density as $V_c\propto 1/\rho$, this model is consistent with the isomorph filter.

\item{{\it Fictive temperature variations following a temperature jump}}

Any jump from equilibrium at some density and temperature to another density and temperature proceeds as if the system first jumped along an isomorph to equilibrium at the final density and then, immediately starting thereafter, jumped to the final temperature (Fig. \ref{fig6}(a)): The first isomorphic jump takes the system instantaneously to equilibrium. This property, which applies for all strongly correlating liquids, means that glass-forming van der Waals and metallic liquids are predicted to have simpler aging behavior than, e.g., covalently bonded liquids like ordinary oxide glasses \cite{gna10}. 

In traditional glass science the concept of ``fictive temperature'' is used as a structural characteristic that by definition gives the temperature at which the structure would be in equilibrium \cite{kov63,moy76,maz77,str78,mck94,hod95}. For any aging experiment, in glass science one assumes that the fictive temperature adjusts itself monotonically from the initial temperature to the final temperature. Consider, however, a sudden temperature increase applied at ambient pressure. In this case there is first a rapid thermal expansion before any relaxation takes place. This ``instantaneous isomorph'' takes the system initially to a state with canonical (Boltzmann) probability factors corresponding to a {\it lower} temperature. In other words, immediately after the temperature up jump the system has a structure which is characteristic of a temperature that is {\it lower} than the initial temperature. With any reasonable definition of the fictive temperature, this quantity thus initially must decrease during an isobaric positive temperature jump -- at least for all strongly correlating liquids. 

\end{enumerate}

\section{Summary}

The class of strongly correlating liquids includes the van der Waals and metallic liquids, but excludes the hydrogen-bonded, the covalently bonded, and the (strongly) ionic liquids. Due to their ``hidden scale invariance'' -- the fact that they inhering a number of IPL properties -- strongly correlating liquids are simpler than liquids in general. Strongly correlating liquids have isomorphs, curves along which a number of physical properties are invariant when given in properly reduced units. In particular, for glass-forming liquids the property of strong virial / potential energy correlations in the equilibrium fluctuations implies a number of experimental predictions. Some of these, like density scaling and isochronal superposition, are well-established experimental facts for van der Waals liquid and known not apply for hydrogen-bonded liquids. This is consistent with our predictions. Some of the predicted properties have not yet been tested, for instance that the density scaling exponent can be determined by measuring the linear thermoviscoelastic response functions at a single state point, or that jumps between isomorphic state points take the system instantaneously to equilibrium, no matter how long is the relaxation time of the liquid at the relevant state points. -- We hope this paper may inspire to new experiments testing the new predictions.

\acknowledgments 
URP  is supported by the Danish Council for Independent Research in Natural Sciences. The centre for viscous liquid dynamics ``Glass and Time'' is sponsored by the Danish National Research Foundation (DNRF).

\end{document}